\documentstyle[11pt] {article}

\title{Distributed Phase Acquisition in a Wave Function}

\author{ R. Englman$^a$ and A. Yahalom $^b$\\
 $^a$ Soreq NRC,Yavne 81800,Israel\\
 $^b$ Ariel University Center of  Samaria, Ariel 40700, Israel\\
e-mail: englman@vms.huji.ac.il; asya@ariel.ac.il;}

\begin{document}
\maketitle

\newcommand{\beq} {\begin{equation}}
\newcommand{\enq} {\end{equation}}
\newcommand{\ber} {\begin {eqnarray}}
\newcommand{\enr} {\end {eqnarray}}
\newcommand{\eq} {equation}
\newcommand{\eqs} {equations }
\newcommand{\mn}  {{\mu \nu}}
\newcommand{\sn}  {{\sigma \nu}}
\newcommand{\rhm}  {{\rho \mu}}
\newcommand{\sr}  {{\sigma \rho}}
\newcommand{\bh}  {{\bar h}}
\newcommand {\er}[1] {equation (\ref{#1}) }
\newcommand {\ern}[1] {equation (\ref{#1})}
\newcommand{\mbf} {{ }}

\begin {abstract}
 A separable $x-y$ model is solved for a specialized vector
 potential (no magnetic and weak electric fields)
 penetrating slowly\textbf{,} adiabatically into and across a rectangular box to which an electron is confined. The
 time-dependent Schr\"odinger equation has  adiabatic solutions, in which gradual
 phase acquisitions occur for {\it parts} of the
 electronic wave function. For a closed trajectory of the source, the initial and after-return wave functions are
 shown to be simultaneously co-degenerate solutions of the  Hamiltonian, which situation repeats itself
  for further cyclic motion of the source.

\end{abstract}

~




Short Title: Distributed Phase Acquisition

~

 In a number of publications,
\cite{BerryW}-\cite {BerryP} and a conference lecture \cite {Popescu1}, the
effect of penetration by a thin solenoid into a confined spatially
extended electron wave function was considered. Clearly, if the solenoid fully
encircles the electron, the vector potential causes an
Aharonov-Bohm (AB) phase factor to get attached to the
wave-function, but a partial penetration into the electron wave function with
eventual return of the solenoid to its starting position has
raised many issues. In the particular case of a semi-fluxon \textbf{,} when
the solenoid intensity is such that the AB phase is just $\pm
\pi$, it was concluded that for one or more positions of the
solenoid inside the electron, energy degeneracies of the
electronic states must occur \cite{Aharonov5, BerryP} and in the
latter work the degeneracy positions were derived for various
shapes of two dimensional confinements. An essential factor in the
derivation by \cite{BerryP} was that a nodal line from the
solenoid to the boundary \textbf{,} or a set of nodal lines is imperative,
so as to avoid phase-caused discontinuities in the wave function.

In this paper we explore by a time-dependent approach and using a
special, idealized model a complementary and new effect. In this model the
penetration into a spatially  extended electronic state is by a
moving classical, macroscopic source giving rise to a vector
potential, that results in  a weak extended electric rather than to
a point-like magnetic field. We find that for a slowly \textbf{,}
adiabatically moving field-source an AB-like phase is engendered
in {\it part} of the extended wave function, together with a
different Berry-like phase for the full wave function. The wave
function part maintains the phase, also after a fully cyclic motion
of the source is completed.

  We consider an electron confined by infinite potential walls $V(x,y)$  to a
   rectangular domain (box) occupying the regions $(0,X)$ in the $x$-direction and
 $(0,Y)$ in the $y$-direction. This might be a drastically simplified two dimensional model for an
 electron attached to its mother nucleus. Stationary energies and eigenfunctions of
the box Hamiltonian are:
\ber
E_{n_x,n_y} & = & \frac{\hbar^2\pi^2}{2m}[(\frac{n_x}{X})^2+(\frac{n_y}{Y})^2]\label{en}\\
f_{n_x,n_y}(x,y) & = & \chi_{n_x}(x)\eta_{n_y}(y)
\label{states}
\enr
with the complete orthonormal sets, appropriate to the fully reflecting \\
boundaries of the confining box, being given by:
\ber
\chi_{n_x}(x)=\sqrt{\frac{2}{X}}\sin \frac{n_x \pi x}{X} \quad x \in (0,X), \qquad \chi_{n_x}(x)=0 \quad {\rm otherwise}
\nonumber \\
~\eta_{n_y}(y)=\sqrt{\frac{2}{Y}}\sin \frac{n_y \pi y}{Y} \quad y \in (0,Y), \qquad \chi_{n_y}(y)=0 \quad {\rm otherwise}.
\label{chieta}
\enr
$n_x,n_y$ are  integers, $m$ is the electronic mass.

 We now introduce the  position dependent "phase function"\textbf{,} which will appear as a phase in the gauge-transformed
 wave function in \er {Psi2} below\textbf{,}
 \beq
 F(x,y,t)=-\Phi\Theta_w(x-vt)\Theta_w(y-Y_s)
 \label{Phasefn}
 \enq
 of strength $\Phi$ and a two-component vector potential $\vec A = \vec \nabla F$ of zero magnetic flux derived from
 it and expressed by:
 \ber
 A_x(x,y,t;v,Y_s,w,\Phi) & = & \frac{\partial }{\partial x}F(x,y,t) =-\Phi \delta_w  (x-vt)\Theta_w(y-Y_s)
 \label{Ax}\\
 A_y(x,y,t;v,Y_s,w,\Phi) & = & \frac{\partial }{\partial y}F(x,y,t) =-\Phi \Theta_w  (x-vt)\delta_w(y-Y_s)
 \label{Ay} \\
 A_z(x,y,t;v,Y_s,w,\Phi) & = & \frac{\partial }{\partial z}F(x,y,t) = 0
 \label{Az}
 \enr
 The functions $\delta_w  (x-vt)$ and $\Theta_w (y-Y_s)$ are broadened modifications of finite but
 arbitrarily small widths,
 of size $2w$ of the well known delta and step functions, used here so as to avoid
 having discontinuities in the wave function. We will also assume  a null $\phi = 0$ scalar potential
  not related to the confining potential.
 The same electromagnetic field can be described by a different set of potentials
 which can be obtained through a gauge transformation
 of the type (in CGS units):
\ber
\vec A_{\Lambda} &=& \vec A + \vec \nabla \Lambda
\nonumber \\
\phi_{\Lambda}&=& \phi - \frac{1}{c} \frac{\partial \Lambda }{\partial t}
\label{gauget}
\enr
Choosing $\Lambda= -F$ we arrive at the Coulomb gauge vector potentials:
\ber
\vec A_C &=& 0
\nonumber \\
\phi_C &=& \frac{1}{c} \frac{\partial F }{\partial t}
\label{gauget2}
\enr
For either choice of gauge we arrive at an electromagnetic field, with zero magnetic field:
\beq
\vec B \equiv curl {\vec A}= 0
 \label{H2}
 \enq
 but with a non vanishing electric field inside the domain:
 \beq
 \vec E \equiv -\vec \nabla \phi -\frac{\partial{\vec A}}{c\partial t} =- \vec \nabla \frac{\partial F}{c\partial t}
 \label{E}
 \enq

The source-charge density $\rho$ and source current density $\vec J$ needed to generate the field are:
\ber
\rho & = & \frac{1}{4 \pi} \vec \nabla \cdot \vec E = - \frac{1}{4 \pi c} \vec \nabla^2 \frac{\partial{F}}{\partial t}
\nonumber \\
\vec J & = & - \frac{1}{4 \pi}\frac{\partial \vec E}{\partial t} = \frac{1}{4 \pi c} \vec \nabla \frac{\partial^2 F}{\partial t^2}
\label{sources}
\enr
Or more explicitly:
\ber
\rho & = & -\frac{v\Phi}{4\pi c}[\delta_w^{''}(x-vt)\Theta_w(y-Y_s)+\delta_w(x-vt)\delta'_w(y-Y_s)]
\nonumber \\
J_x & = & -\frac{v^2\Phi}{4\pi c} \delta_w^{''}(x-vt)\Theta_w(y-Y_s)
\nonumber \\
J_y & = & -\frac{v^2\Phi}{4\pi c} \delta_w^{'}(x-vt)\delta_w(y-Y_s)
\nonumber \\
J_z & = & 0
\label{sources2}
\enr
where an apostrophe denotes differentiation with respect to the argument.

 The field is thus due to a wire and sheet sources\textbf{,} infinitely long in the z-direction\textbf{,} as they slowly sweep across
 the domain with a "horizontal" velocity $v$ coming
 from the right ($v<0$) and mainly operating in the region of $y$-values greater or equal to
 $Y_s$\textbf{,} give and take a width $w$, much smaller than the domain dimensions. The source motion starts
 at negative times at the far right, reaches
 the box at $t=-X/|v|$, exits it at $t=0$ and is allowed to make a return above
 the box  to its initial position\textbf{.} This is not described by the potential of \ern{Phasefn}. The latter part of the
 trajectory is of no immediate interest here, as it does not affect the wave function, but will enter later, when we
 regard the source motion as periodic \cite{squaremotion}.
 Due to locality\textbf{,} the vector potential only interacts with the wave function
 when the source penetrates the  box. Its non-zero parts and that of the
 phase function $ F(x,y,t) $ are depicted in Figure \ref{fig1}.
\figure
 \vspace{14cm}
 \includegraphics{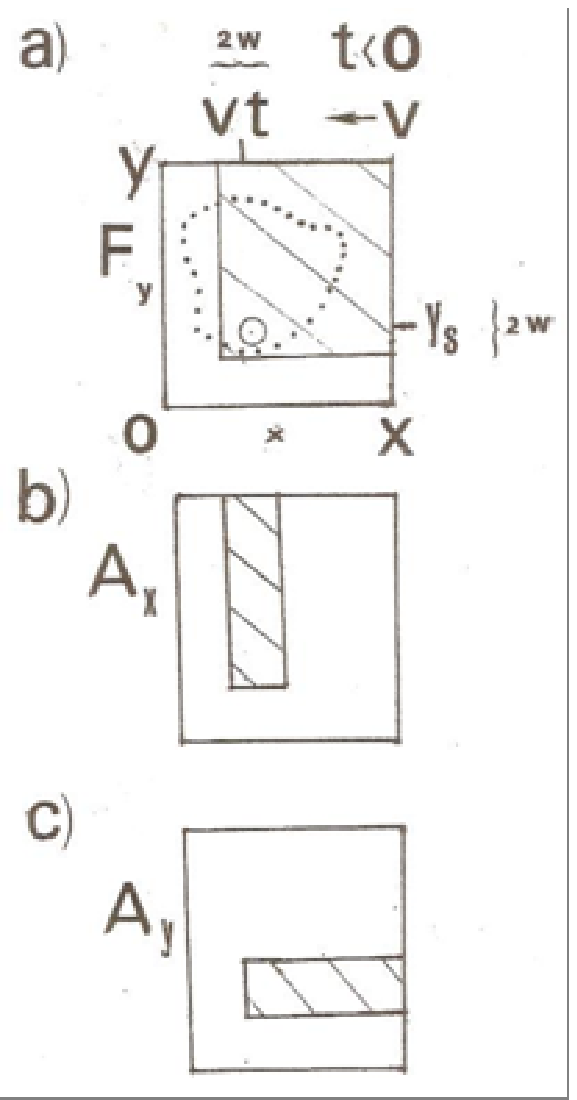}
\caption{ Non-zero regions of the functions appearing in this paper: shown
 by shading.
(a) Regions of the phase function $F(x,y,t)$\textbf{,} \ern{Phasefn}. The
tip of the source indicated by a circle and moving with a velocity $v (<0)$
is inside the  $XY$ rectangle  at times $X/v\leq t\leq 0$ and enters the
box at a height of $y=Y_s$. $2w$ indicates schematically the
function broadening. A domain with arbitrary shape is drawn with dotted lines.
(b) Non-zero regions of the vector potential component $A_x$.
(c) Same for $A_y$.}
\label{fig1}
\endfigure

 A  vector potential functionally similar to that above,  was used in  \cite {AharonovK}.
  The gauge dependence of $\vec A$ is  nullified (for single valued gauges) by letting the source
  return to its initial position \textbf{,}
  rather than moving along a straight line\textbf{,}
 and considering the electronic state after return of the source to its starting position.
 In the present case an arbitrary gauge
 transformation can be applied to the Hamiltonian\textbf{,} together with a corresponding gauge factor in the
 wave function\textbf{,} without
 affecting the end result. In particular, one can work with a Coulomb
 gauge\textbf{,} \ern{gauget2}. The reason of
 our preference for the vector potential description is that the references in the Introduction used this gauge\textbf{,}
 although with different forms for ${\vec A}$.

 The present model takes us to a Hamiltonian:
 \beq
 H(x,y,t)= \frac{1}{2m}[(p_x-\frac{e}{c}A_x(Y_s,t))^2+(p_y-\frac{e}{c}A_y(Y_s,t))^2] +  V(x,y)
\label{H}
\enq
where $V(x,y)$ is the potential confining the particle inside the rectangle of size $X-Y$.

  Since the Hamiltonian is time dependent, electronic states
 $\Psi(x,y,t)$
 will be obtained from the time dependent Schr\"odinger equation.
 \beq
 i\hbar \frac{\partial\Psi(x,y,t)}{\partial
 t}=H(x,y,t)\Psi(x,y,t)
 \label {SE}
 \enq
 We consider a solution whose initial form is fixed at the instant $t=X/v$, \textbf{this} being negative since $v<0$.

  \beq
  \Psi(x,y,t \simeq X/v) =\chi_{n_x}(x)\eta_{n_y}(y)
 \label{start}
 \enq
 (cf. \ern{chieta}.) We now postulate the  form for the state evolving adiabatically:
\beq
\Psi_{n_x n_y}(x,y,t)=e^{iQ(x,y,t)}\chi_{n_x}(x)\eta_{n_y}(y)
 \label{Psi2}
 \enq
 \textbf{where \beq e^{iQ(x,y,t)}=e^{-iE_{n_x n_y}t/\hbar}e^{i\gamma_{n_x
 n_y}(t)}e^{i\frac{e}{\hbar  c}F(x,y,t)}\label{Q}\enq}
  As will emerge, the first factor is the dynamic phase factor (here appearing in  a
 trivial form), the second is a geometric phase factor and the third
 factor is  chosen, following \cite{BerryP}, in order to eliminate the vector potential and the time dependence from
 the Hamiltonian through a gauge-transformation $U$ where:
 \beq  U \equiv  U(x,y,t)  =  e^{i\frac{e}{\hbar c}F(x,y,t)}
 \label{U}
 \enq
 so that:
 \ber
 U^{-1}H(x,y,t)U \psi_0 (x,y,t)  &=&
\nonumber \\
 H(x,y) \psi_0 (x,y,t) & \equiv &  \left[ \frac{1}{2m}[p_x^2+p_y^2] + V(x,y)\right]\psi_0 (x,y,t)
\label{U2}
\enr
for any wave function $\psi_0 (x,y,t)$. The $H(x,y)$
eigen functions are clearly of the form shown in \ern{chieta}.
Substituting $\Psi(x,y,t)$ from \er{Psi2} into the  Schr\"odinger
\er{SE} and multiplying through with $ e^{-iQ(x,y,t)}$\textbf{one obtains}
\ber e^{-iQ(x,y,t)}i\hbar \frac{\partial [e^{iQ(x,y,t)}\chi_{n_x}(x)\eta_{n_y}(y)]}{\partial
 t} & = & E_{n_x n_y}\chi_{n_x}(x)\eta_{n_y}(y)+ \delta_{n_x n_y}(x,y,t)\nonumber\\
    & = &H(x,y)\chi_{n_x}(x)\eta_{n_y}(y)\label{RE}\enr
 \textbf{Thus, arising from the time derivative one has the leftover term}
 \beq \delta_{n_x n_y}(x,y,t) =
  \left[-\hbar{\dot\gamma_{n_x  n_y}(t)}- \frac{e}{c} \dot F(x,y,t) \right] \chi_{n_x}(x)\eta_{n_y}(y)
 \label {leftover}
 \enq
where the dot signifies time differentiation.

Expanding this expression in terms of the complete set
$\chi_{m_x}(x)\eta_{m_y}(y)$, one sees that the first term in \er{leftover} has
only "diagonal" components ($m_x=n_x,m_y=n_y$), while the second
has both diagonal components\textbf{,} which will give the Berry phase
\cite {Berry}\textbf{,} and non-diagonal components, which will be shown to
be small in an adiabatic motion of the source.

Dealing first with the latter, we have for the absolute magnitude
of the expansion coefficients arising from the second term only
\ber
&&\hspace{-5mm}|\frac{ev \Phi}{c}\int_0^X dx
\chi_{m_x}(x)\chi_{n_x}(x)\delta_s(x-vt)\int_0^Y dy
\eta_{m_y}(y)\eta_{n_y}(y)\Theta_s(y-Y_s)| \leq
 \nonumber\\
&&\hspace{-5mm} |\frac{ev \Phi}{c}|\int_0^X dx
|\chi_{m_x}(x)\chi_{n_x}(x)|\delta_s(x-vt)\int_0^Y dy
|\eta_{m_y}(y)\eta_{n_y}(y)|\Theta_s(y-Y_s)  \leq
\nonumber\\
&&\hspace{-5mm} |\frac{ev \Phi}{c}|[{\frac{2}{X}\int_0^X dx \delta_s(x-vt)}]\cdot
[\frac{2}{Y}\int_0^Y dy |\Theta_s(y-Y_s)]  \leq
\nonumber\\
&&\hspace{-5mm} |\frac{4 e v \Phi}{cX}|\cdot \frac{Y-Y_s}{Y}
\label{nondiag}
\enr
for the non-trivial case that the source is inside the box ($Y_s\leq Y$). In going from the second
to the third line we have inserted the forms of the functions from
\er{chieta} and also noted that the sines are equal or less than
unity. For the ${m_x m_y -n_x n_y}$  non-diagonal expansion
coefficients to be negligible, these must be much less than the
corresponding energy change, or
\beq
|\frac{4 e v \Phi}{c X}|\cdot \frac{Y-Y_s}{Y}<< electronic ~energy~ differences =O(E_{n_xn_y})|
\label{criterion}
\enq
which is clearly satisfied in the  adiabatic limit, with
sufficiently low velocities $|v|$.

Turning to the diagonal part of the leftover terms, this vanishes
provided:
\beq
{\dot\gamma_{n_x n_y}(t)}=i<n_x n_y,t|\frac{\partial}{\partial t}|n_x n_y,t>
\label{gdot}
\enq
where the Dirac ket represents:
\beq
|n_x n_y,t> \equiv e^{i\frac{e}{\hbar c}F(x,y,t)} \chi_{n_x}(x)\eta_{n_y}(y)
\enq
 $\gamma(t)$ the time integral of this, is an overall
phase of the wave function for all values of $t$ and, when $t$ is
not a period of the source motion, is the "Open path phase" \cite
{Bhandari}. As noted before, for $t=\tau$ the period in a cyclic
motion of the source, $\gamma(\tau)$ is identified with the Berry
phase. In the present context, it is a function of the vertical
entrance position $Y_s$ (which appears in the phase function
$F(x,y,t)$ of \ern{Phasefn}). The Berry phase {\it tends} to cancel
the acquired partial phases in the wave function, but achieves
full cancellation only when the source motion fully encompasses
the box \cite{ReznikA}. To see this we elaborate on the
full-period time integral of the right hand side of the above
equation, as follows (dispensing, as we go along, with the wave
function indices):
\ber
 && i\oint dt <n_x n_y,t|\frac{\partial}{\partial t}|n_x n_y,t>
 \nonumber \\
&=& - \frac{ev\Phi}{\hbar c}\int_{X/v}^0 dt\int_0^X dx [\chi(x)]^2 \int_{0}^Y dy[\eta(y)]^2 \Theta_s(y-Y_s)\delta_s(x-vt)
\label{BP2}\\
& = & -\frac{ev\Phi}{\hbar c}\int_{X/v}^0 dt[\chi(vt)]^2\int_{Y_s}^Y dy[\eta(y)]^2
\label{BP3}\\
& = & \frac{e\Phi}{\hbar c}\int_{0}^X d(vt)[\chi(vt)]^2\int_{Y_s}^Y dy [\eta(y)]^2
\label{BP3b}
\enr
The equality after the first line is
based on the fact that the chosen set vanishes outside the box. In
\er {BP3} the first integration\textbf{,} over $vt$\textbf{,} covers the full $x-$
range of the confining box and therefore gives just $1$, but the
second integration\textbf{,} over the truncated $y$-range\textbf{,} will be less
than $1$, unless $Y_s\leq 0$, meaning that the source passes at
the bottom edge of the box, or below. Thus \er {BP3b} is generally
less than the maximal AB phase. On the other hand, in the
phase $\frac{e}{\hbar c}F(x,y,t)$ of the wave function, the part
in which  $x>|vt|$ and $y>Y_s$ (namely, above the source-motion
line) gets the full AB phase and that in which $x<|vt|$ or $y<Y_s$
(below the source-motion line)  gets no phase. Thus the Berry phase
does not in general cancel the partial phases.

 The solution to \ern{SE}, with the proclaimed
 initial condition and insertion of the open path phase from the previous
   section, is that shown in \ern{Psi2}.
In the limit of an infinitesimally narrow source  $(w\to 0)$, this
wave function without the dynamic and open path factors has the
following meaning:

The initial state is kept for source positions "above" the box
($Y_s>Y$)  or before reaching the box, but a phase is acquired if
the source penetrates or is "under" the box and then also only for
that $x$-part (vertical section) of the wave function which the
source has already passed. The overall phases which
the adiabatic state acquires are the dynamic phase $-E_{n_x
n_y}t/\hbar$ and the open path or Berry phase $\gamma_{n_x
 n_y}(t)$. There is no
degeneracy in the system and, indeed, the energy $E_{n_x n_y}$ of
the state not only does not coalesce with a neighboring one, but
even stays constant during the source motion: an exceptional
situation for adiabatic or any type of motion.

 The described phase acquisition process \textbf{,} with the function-widths $w$ non-zero\textbf{,}
 is entirely continuous
  in  that the region in which the new phase appears grows steadily from zero as
 the source enters the box from the right or from above and ultimately  covers the whole
 box, which situation persists as the source passes the box from below.

 Objection may be raised that in an adiabatic  motion with a
periodic Hamiltonian, at the completion of a period, the adiabatic
wave function must return to its initial form except for an
overall phase, whereas this is not the case for the adiabatic
solution in \er {Psi2}. However, this objection
holds only for states that are not simultaneous eigenstates of the
Hamiltonian, whereas the following proof shows that in the present
case the wave functions obtained after an integral number of
periods, although different, are co-degenerate solutions of the
{\it same} Hamiltonian.

~

Proof (In the proof the coordinate dependence of the functions, as
well as any {\it overall} time-dependent phase factor, are
omitted, for brevity):

$\Psi_{n_xn_y}(t)= U(t)\chi_{n_x}\eta_{n_y} $ solves the periodic
Hamiltonian $H(t)=H(t+\tau)$\textbf{,} $\tau$ being the period\textbf{,} with an
eigenvalue $E_{n_xn_y}$ since
\ber H(t)\Psi_{n_xn_y}(t) & = &
U(t)U^{-1}(t)H(t)U(t)\chi_{n_x}\eta_{n_y} = U(t)H
\chi_{n_x}\eta_{n_y} \nonumber
\\ & = & U(t) E_{n_xn_y}\chi_{n_x}\eta_{n_y} =
E_{n_xn_y}\Psi_{n_xn_y}(t)\label{soln1}\enr where $H$ denotes the
time independent Hamiltonian in \ern{U2}. We now show that the wave
function $\Psi_{n_xn_y}(t+\tau)$ one period later\textbf{,} is also a
solution with the same eigenvalue of the {\it same} Hamiltonian
$H(t)$. (For $\tau$ not a period, the later wave function is a
solution with the same eigenvalue, however, not of $H(t)$ but of a
{\it different} Hamiltonian $H(t+\tau)$.) \ber
H(t)\Psi_{n_xn_y}(t+\tau) & = &
U(t+\tau)U^{-1}(t+\tau)H(t)U(t+\tau)\chi_{n_x}\eta_{n_y} \nonumber\\
 & = & U(t+\tau)U^{-1}(t+\tau)H(t+\tau)U(t+\tau)\chi_{n_x}\eta_{n_y}\nonumber\\
 & = & U(t+\tau)H \chi_{n_x}\eta_{n_y} \nonumber \\ & = & U(t+\tau)
E_{n_xn_y}\chi_{n_x}\eta_{n_y} =
E_{n_xn_y}\Psi_{n_xn_y}(t+\tau)\nonumber\\ & & \label{soln2}\enr
where in going from the first to the second line we have used the
periodicity of the Hamiltonian.

Thus, after one full cycle the wave function turns, instead of the
initial state, into a co-degenerate state, different from but not
orthogonal to the initial one. If the AB phase is a rational
submultiple of $2\pi$, each further cycle will bring the system to
a new co-degenerate state, until the accumulated phases add up to
$2\pi$ (or its integral multiple), i.e. the initial state is
regained.

 It should also be noted that this
"periodicity-degeneracy" involving all wave functions differing
by a multiple of the trajectory period is unrelated to the
degeneracies found at discrete values of the solenoid coordinates
in \cite{BerryP}.

In conclusion, we have considered the adiabatic penetration into a spatially
extended electron  by a source, which is such that there is no
magnetic field, only a weak, motion-induced electric field. A distributed (i.e., not overall)
phase acquisition by parts of the electron occurs, and agglomerates in
the course of continued cyclic motion. A remarkable feature of
this outcome is that whereas \textbf{,} in the limit of extreme
adiabaticity\textbf{,} the electric field is too weak to cause energy
excitations, the vector potential does change the phases
(differentially and locally). This resembles some analogous
features in the AB and Berry's phase effects (no
fields and still phase change). Indeed the present result can be
formulated as a weakened AB effect, in that a
measurable phase acquisition by the vector potential alone occurs
in regions where there is no magnetic field and the electric
field, though not nil, is so weak that it causes no excitation or
admixture.

The novelty in this letter's time dependent treatment\textbf{,} compared to \cite{BerryW}-\cite{BerryP}\textbf{,}
is the absence of local degeneracies and the phase
acquisition by parts of the wave packet. These results carry over to generalizations
regarding  forms of the
confining box illustrated by dotted lines in Figure 1 (a) and of vector
 potentials derived from arbitrary forms of the phase functions,
  subjects that will be treated in the future. Future study will also consider many
  particle effects, resulting in specific braiding properties\textbf{,} e.g.\textbf{,} the acquisition
  by the many particle wave-function of a {\it non}-integral $\pi$ phase upon
  adiabatic interchange of two neighboring particles.

\end {document}